\documentclass[aps,pra]{revtex4}
\usepackage{times}
\usepackage{epsfig}
\usepackage{amssymb}
\usepackage{amsmath}
\usepackage{amsfonts}
\usepackage{bm}

\begin{document}
\title{Radiative QED corrections to one-photon transition rates in hydrogen atom at finite temperatures}
\author{T. Zalialiutdinov$^{1}$, D. Solovyev$^1$ and L. Labzowsky$^{1,2}$, }

\affiliation{ 
$^1$ Department of Physics, St. Petersburg
State University, Petrodvorets, Oulianovskaya 1, 198504,
St. Petersburg, Russia
\\
$^2$ Petersburg Nuclear Physics Institute, 188300, Gatchina, St. Petersburg, Russia}

\begin{abstract}
Within the framework of QED theory at finite temperature the thermal radiative corrections to spontaneous one-photon transition rates in hydrogen atom are investigated. The radiative one-loop self-energy corrections are described in the thermal case. Closed analytical expressions are derived and their numerical calculations for the spontaneous decay rate of the $ 2p $ state are carried out. Dominance of thermal radiative corrections to spontaneous  Ly$_{\alpha}  $ decay rate over ordinary induced transition rate up to temperatures $ T  < 6000 $ K is demonstrated.
\end{abstract}

\maketitle
\section{Introduction}
\label{sectionINTRO}

Radiative corrections to transition rates and lifetimes in atoms and ions are of no less interest than radiative corrections to the energies of bound states (such as corrections to the Lamb shift, hyperfine splitting etc.) \cite{drake,sucher,saperstein,volotka}. The precision in measurements of decay rates has considerably increased in recent years which has been accompanied by a corresponding increase in the accuracy of theoretical calculations \cite{st1,st2,st3,st4,st5,st6,st7}. The experimental uncertainty achieved at the level of one per thousand \cite{tupvol2} makes such studies sensitive to relativistic, radiative, nuclear size and  recoil effects \cite{rec1,rec2,rec3,tup,volotka,drakerel}. Therefore, the detailed theoretical analysis of various radiative corrections providing a versatile verification of fundamental physics is required. The precise values of the transition rates in different atomic systems are also needed for investigations of atomic collision processes or interpretation of the spectra from astrophysical sources \cite{source}. Moreover, accurate calculations of the transition rates can serve for verifications of basic parts in more complicated processes, such as, e.g., the parity violation amplitudes in heavy ions and atoms \cite{saper2,shab0}. In this regard, radiative corrections to the transition rates become extremely important for the decay rates suppressed with respect to dipole transitions (forbidden by selection rules) \cite{drake, tupvol1}. 
To date, theoretical calculations of such decay rates had advanced to evaluation of the two-loop self-energy diagrams \cite{saperstein, zalialiutdinov, jent}. 

The accounting of different radiative effects required to achieve the experimental uncertainty draws attention to the phenomena of other type. The impact of physical conditions, such as external fields for example, on laboratory experiment and astrophysical processes deserves a special consideration \cite{riehle,labbook,fradkin,kaplan,sol-ext}. As a separate area of ​​research, the influence of blackbody radiation (BBR) at finite temperatures plays an important role in a number of scenarios: development of atomic clocks, recombination history of early universe or radiation transfer in interstellar medium \cite{safronova1, middelmann, safronova2, beterov, ovsiannikov, chluba, hirata}. The quantum mechanical (QM) theory of energy shift and transition rates induced by the BBR was given in \cite{farley}. 
Theoretical calculations within the framework of rigorous quantum electrodynamics (QED) of the thermal Stark shift, level broadening and BBR-induced bound-free transitions have been carried out in \cite{solovyev2015}, and later in \cite{zalialiutdinov2018,zalialitdinov2019, solovyev2019}. The advantage of QED theory application to investigations of this kind is the accurate accounting for finite life times of atomic levels. Although such effects are outside the scope of this work, the formalism developed in \cite{solovyev2015,zalialiutdinov2018,zalialitdinov2019, solovyev2019}, see also \cite{solovyevarxiv2019}, can reveal new (unknown) thermal corrections to transition rates between bound states and to analyze their significance at astrophysical and laboratory conditions.
 
In the present work the radiative one-loop self-energy (SE) corrections caused by the "heat bath" to the one-photon transition rates for the hydrogen atom are investigated within the framework of thermal QED theory. The heat bath acting on the atomic system implies an environment described by blackbody radiation, i.e. the photon field distributed according to Planck's law. We restrict ourselves to considering the leading SE thermal corrections to the transition rates, since the next orders or effects associated with vacuum polarization (VP) are suppressed by temperature factors and, therefore, should be much less \cite{drake,volotka,solovyevarxiv2019}.

The paper is organized as follows. In section \ref{sectionA} we briefly describe adiabatic S-matrix approach for the evaluation of one-photon transition probabilities in one-electron atomic systems. The derivation of SE thermal radiative corrections is considered in section \ref{sectionB}. Expressions derived in section \ref{sectionB} are applied to the numerical calculation of thermal corrections to $ 2p\rightarrow 1s+\gamma(\mathrm{E1}) $ transition rate in hydrogen. The results of calculations are discussed in section \ref{theend}. Below we will use the relativistic units $ \hbar=m_{e}=c=1 $ ($ m_{e} $ is the electron rest mass, $c$ is the speed of light and $\hbar$ is the reduced Planck constant).

\section{Adiabatic S-matrix approach: evaluation of transition rate}
\label{sectionA}

For evaluation of the transition probabilities and radiative corrections we will use the adiabatic S-matrix approach \cite{low, olegreports,physrep2018}. This allows one to take into account contributions to QED corrections arising from the reducible Feynman diagrams \cite{lab93,labbook}. The adiabatic S$_{\eta} $-matrix differs from the ordinary S-matrix by the insertion of exponential factor $ e^{-\eta |t|} $ ($ \eta>0 $ is the adiabatic parameter) in each vertex of Feynman diagram. It refers to the concept of adiabatic switching on and off the interaction introduced formally by the replacement $\hat{H}_{\mathrm{int}}(t) \rightarrow \hat{H}^{\eta}_{\mathrm{int}}(t) = e^{-\eta|t|}\hat{H}_{\mathrm{int}}(t)$ \cite{sucher, saperstein}.

\begin{figure}
\centering
\includegraphics[scale=0.7]{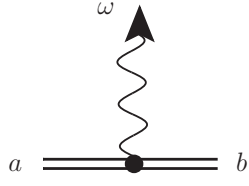}
\caption{The Feynman graph corresponding to the photon emission in an one-electron atom. The double solid line describes the electron in field of nucleus (Furry picture), the wavy line with arrow at the end describes emitted photon. The indices $a$ and $ b $ refer to the set of quantum numbers of the initial and final states of an atom, respectively, $ \omega $ denotes the frequency of emitted photon.}
\label{fig1}
\end{figure}

For one-electron atom the one-photon transition from the state $ a $ to the state $ b $ is described by the Feynman diagram in Fig. \ref{fig1}. Within the framework of adiabatic S-matrix approach the corresponding first order S-matrix element is
\begin{eqnarray}
\label{1a}
\hat{S}^{(1)}_{ab}=(-\mathrm{i}e)\int dx \overline{\psi}_{b}(x)\gamma_{\mu}A^{*}_{\mu}(x)e^{-\eta |t|}\psi_{a}(x)
.
\end{eqnarray}
Here $ e $ is the electron charge, $ \psi_{a}(x)=\psi_{a}(\textbf{r})e^{-\mathrm{i}E_{a}t} $ is the solution of Dirac's equation for the atomic electron, $ E_{a} $ is the Dirac energy of the state $ a $, $ \overline{\psi}_{a}=\psi_{a}^{\dagger}\gamma_{0} $ is the Dirac conjugated wave function with $ \psi_{a}^{\dagger} $ being its Hermitian conjugate and $ \gamma_{\mu}=(\gamma_{0},\bm{\gamma}) $ are the Dirac matrices. Photon field wave function $ A^{(\textbf{k},\textbf{e})}_{\mu}(x) $ is
\begin{eqnarray}
\label{2a}
A^{(\textbf{k},\textbf{e})}_{\mu}(x)=\sqrt{\frac{2\pi}{\omega}}e^{(\lambda)}_{\mu}e^{\mathrm{i}(\textbf{k}\textbf{r}-\omega t)}=\sqrt{\frac{2\pi}{\omega}}e^{(\lambda)}_{\mu}e^{-\mathrm{i}\omega t}A^{(\textbf{k},\textbf{e})}_{\mu}(\textbf{r})
,
\end{eqnarray}
where $  e^{(\lambda)}_{\mu} $ is the polarization 4-vector, $\textbf{k}$ is the wave vector, $\omega = |\textbf{k}|$ is the photon frequency, $ x\equiv(\textbf{r},t) $ is the coordinate 4-vector, where ($\textbf{r}$, $t$ are the space- and time-coordinates). 

Following the standard evaluation of S-matrix theory the transition rate is \cite{olegreports,physrep2018}
\begin{eqnarray}
\label{4a}
W_{ab}=\lim\limits_{\eta\rightarrow 0+}\eta\sum_{\textbf{e}}\int|\hat{S}^{(1)}_{ab}|^2\frac{d\textbf{k}}{(2\pi)^3}
,
\end{eqnarray}
where $ d\textbf{k}=\omega^{2}d\omega d\bm{\nu} $ and $ \bm{\nu}=\textbf{k}/|\textbf{k}| $ is the photon propagation vector. The integration over time variable in Eq. (\ref{1a}) yields essentially a representation of the $ \delta $-function \cite{ambiguity, physrep2018}
\begin{eqnarray}
\label{3a}
\int\limits_{-\infty}^{\infty}dt e^{\mathrm{i}(E_{b}-E_{a}+\omega)t-\eta |t|}=\frac{2\eta}{(\omega_{ab}-\omega)^2+\eta^2}\equiv 2\pi\delta_{\eta}(\omega_{ab}-\omega)
.
\end{eqnarray}
Here $ \omega_{ab}=E_{a}-E_{b} $ and $ \lim\limits_{\eta\rightarrow 0+}\delta_{\eta}(x)=\delta(x) $.
Then, taking Eq. (\ref{1a}) by square modulus and integrating over $ \omega $, one can arrive at
\begin{eqnarray}
\label{5a}
W_{ab}=\frac{\omega_{ab}^2}{(2\pi)^2}\sum\limits_{\textbf{e}}\int |\hat{U}_{ab}^{(1)}|^2 d\bm{\nu}
,
\end{eqnarray}
where
\begin{eqnarray}
\label{6a}
\hat{U}_{ab}^{(1)}=(-\mathrm{i}e)\sqrt{\frac{2\pi}{\omega_{ab}}}\langle b| (\textbf{e}^{*}\bm{\alpha})e^{-\mathrm{i}\textbf{k}\textbf{r}} |a\rangle
.
\end{eqnarray}
To obtain Eq. (\ref{5a}), the following relation was used \cite{ambiguity,physrep2018}:
\begin{eqnarray}
\label{7a}
4\eta^2 \int\limits_{-\infty}^{+\infty}\frac{\omega d\omega}{((\omega_{ab}-\omega)^2+\eta^2)^2}=\frac{2\pi\omega_{ab}}{\eta}
.
\end{eqnarray}

In the nonrelativistic limit, $ \textbf{k}\textbf{r}\sim \alpha Z \ll 1 $, the use of the dipole approximation for the transition amplitude (\ref{6a}) leads to
\begin{eqnarray}
\label{8a}
\hat{U}^{(1)}_{ab}=(-\mathrm{i}e)\sqrt{\frac{2\pi}{\omega_{ab}}}\langle b |\textbf{e}^{*}\textbf{p}|a\rangle
,
\end{eqnarray}
where $ \langle a|\hat{T} |b\rangle  $ denotes now the matrix element of operator $ \hat{T} $ with Schr\"odinger wave functions for atomic electron in the Coulomb field. Substituting Eq. (\ref{8a}) into Eq. (\ref{5a}), performing summation over photon polarizations and integration over photon directions one can arrive at
\begin{eqnarray}
\label{9a}
W_{ab}=\frac{4e^2}{3}\omega_{ab}|\langle b |\textbf{p}|a\rangle|^2.
\end{eqnarray}
Finally, the summation over magnetic quantum numbers of the final state and averaging over magnetic quantum numbers of the initial state in Eq. (\ref{9a}) should be performed. Then, employing the quantum mechanical relation 
\begin{eqnarray}
\label{relation}
\langle a|\textbf{p}| b \rangle=\mathrm{i}\omega_{ab}\langle a |\textbf{r}| b \rangle
,
\end{eqnarray}
the partial transition rate corresponding to the emission process $a\rightarrow b+\gamma(\mathrm{E1})$ is given by the expression:
\begin{eqnarray}
\label{9aa}
W_{ab}=\frac{4e^2}{3}\frac{1}{2l_{a}+1}\sum\limits_{m_am_b}\omega_{ab}^3|\langle b |\textbf{r}|a\rangle|^2
.
\end{eqnarray} 

In presence of the BBR field, i.e. isotropic external radiation field with equilibrium temperature $T$, the partial induced transition can be additionally found in the form:
\begin{eqnarray}
\label{10a}
W^{\mathrm{ind}}_{ab}=\frac{4e^2}{3}\frac{1}{2l_{a}+1}\sum\limits_{m_am_b}\omega_{ab}^3|\langle b |\textbf{r}|a\rangle|^2 n_{\beta}(\omega_{ab})
,
\end{eqnarray}
where $n_\beta(\omega)$ is the Planck distribution function: $n_\beta(\omega)  = \left(e^{\beta\omega}-1\right)^{-1}$, $\beta\equiv 1/k_B T$, $k_B$ is the Boltzmann constant. 
Thermal background leads to the level broadening due to the BBR induced transition to all possible final states. This broadening is represented by the sum over all final states (including continuum) \cite{farley}:
\begin{eqnarray}
\label{11a}
\Gamma^{\mathrm{BBR}}_{a}=\frac{4e^2}{3}\frac{1}{2l_{a}+1}\sum\limits_{b}\sum\limits_{m_am_b}\omega_{ab}^3|\langle b |\textbf{r}|a\rangle|^2 n_{\beta}(\omega_{ab})
.
\end{eqnarray}
Then the total widths of the level $ a $ is
\begin{eqnarray}
\label{12a}
\Gamma_{a} = \sum\limits_{b<a}W_{ab} + \sum\limits_b W^{\mathrm{ind}}_{ab} \equiv \Gamma^{\mathrm{nat}}_{a}+\Gamma^{\mathrm{BBR}}_{a},
\end{eqnarray}
where $ \Gamma^{\mathrm{nat}}_{a} $ is the natural width of the level $ a $. In the laboratory experiments $\Gamma^{\mathrm{BBR}}_{a}$ is actually small for the low-lying atomic levels but becomes important for the Rydberg states and increases with the growth of temperature. The induced transition rates with necessity are taken into account in the astrophysical investigations at high temperatures \cite{chluba}.

\section{Thermal self-energy corrections to one-photon transition rate}
\label{sectionB}

\begin{figure}[hbtp]
\centering
\includegraphics[scale=0.7]{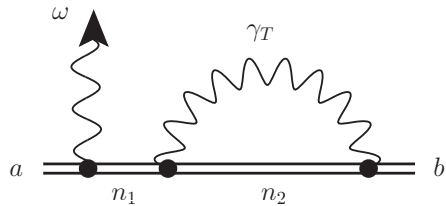}
\caption{The Feynman graph corresponding to thermal one-loop QED correction to photon emission in an one-electron atom. The double solid line describes electron in the field of nucleus (Furry picture), the wavy line with arrow at the end describes emitted photon with frequency $ \omega $. The wavy line with index $ \gamma_{T} $ denotes thermal photon propagator. The indices $  a$ and $ b $ refer to the quantum numbers of the initial and final states of an atom, respectively, while indices $ n_1 $ and $ n_2 $ refer to quantum numbers of intermediate states in electron propagators. }
\label{fig2}
\end{figure}

\begin{figure}[hbtp]
\centering
\includegraphics[scale=0.7]{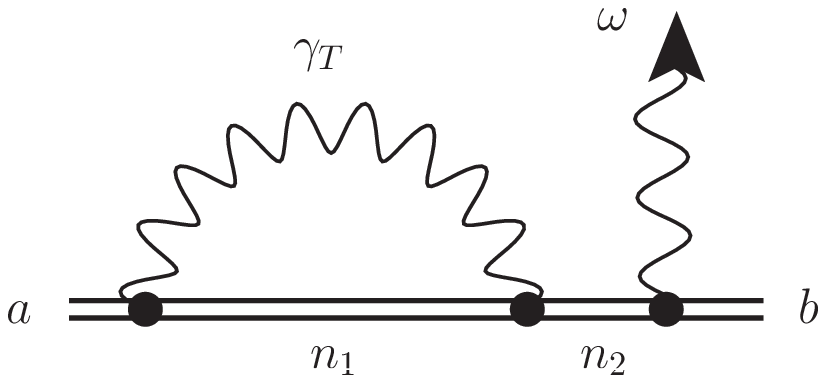}
\caption{The Feynman graph corresponding to the thermal one-loop QED correction to photon emission in an one-electron atom. All notations are the same as in Fig. \ref{fig2}.}
\label{fig3}
\end{figure}

\begin{figure}[hbtp]
\centering
\includegraphics[scale=0.7]{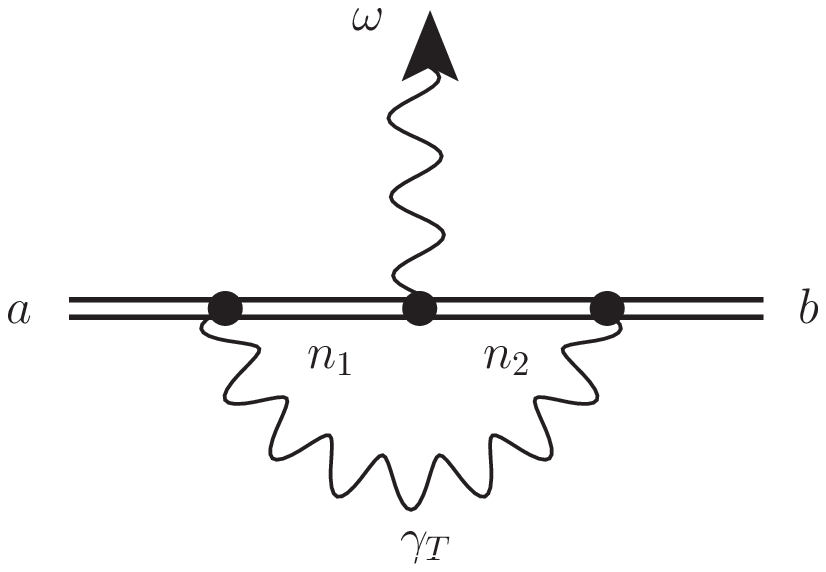}
\caption{The Feynman graph corresponding to the thermal one-loop QED correction to photon emission in an one-electron atom. All notations are the same as in Fig. \ref{fig2}.}
\label{fig4}
\end{figure}

Thermal (one-loop) self-energy corrections to one-photon transition rate are given by the set of Feynman diagrams represented in Figs. \ref{fig2}-\ref{fig4}.
Then the one-photon transition probability corrected according to Figs.~ \ref{fig2}-\ref{fig4} is defined by
\begin{eqnarray}
\label{1b}
\widetilde{W}^{}_{ab}=\lim\limits_{\eta\rightarrow 0+}\eta\sum_{\textbf{e}}\int|\hat{S}^{(1)}_{ab}+\hat{S}^{(3)}_{ab}|^2\frac{d\textbf{k}}{(2\pi)^3}
,
\end{eqnarray}
where $\hat{S}^{(3)}_{ab}$ is the sum of three third-order S-matrix elements corresponding to the Feynman diagrams in Figs. \ref{fig2}-\ref{fig3}. Taking the square modulus in Eq. (\ref{1b}) and neglecting by the terms proportional to $ e^6 $ the corrected transition rate takes the form:
\begin{eqnarray}
\label{2b}
\widetilde{W}_{ab}=\lim\limits_{\eta\rightarrow 0+}\eta\sum_{\textbf{e}}\int\left(|\hat{S}^{(1)}_{ab}|^2+2\mathrm{Re}(\hat{S}^{(1)*}_{ab}\times\hat{S}^{(3)}_{ab})\right)\frac{d\textbf{k}}{(2\pi)^3}
,
\end{eqnarray}
where the first term in brackets in Eq. (\ref{2b}) leads to the expression (\ref{5a}) and the second term represents the thermal corrections. 

We start from the evaluation of thermal loop correction to one-photon transition rate given by Fig. \ref{fig2}. The corresponding adiabatic S-matrix element is
\begin{eqnarray}
\label{3b}
\hat{S}^{(3)\mathrm{Fig.2}}_{ab}=(-\mathrm{i}e)^3\int dx_{3}dx_{2}dx_{1}\overline{\psi}_{b}(x_3)S(x_3,x_2)e^{-\eta |t_3|}\gamma_{\mu_3}D^{\beta}_{\mu_3\mu_2}(x_3,x_2)e^{-\eta |t_2|}\gamma_{\mu_2}S(x_2,x_1)\gamma_{\mu_1}e^{-\eta |t_1|}
A^{*}_{\mu_1}(x_1)\psi_{a}(x_1)
,
\end{eqnarray}
where $  S(x_2x_1)$ is the Feynman propagator for the atomic electron and $ D^{\beta}_{\mu\nu}(x_2,x_1) $ is the thermal photon propagator. In Furry picture the eigenmodes decomposition of the electron propagator reads \cite{akhiezer}
\begin{eqnarray}
\label{4b}
S(x_2,x_1)=\frac{1}{2\pi \mathrm{i}}\int\limits_{-\infty}^{\infty}d\Omega\, e^{\mathrm{i}\Omega(t_2-t_1)} \sum_{n}\frac{\psi_{n}(\textbf{r}_2)\overline{\psi}_n(\textbf{r}_1)}{E_{n}(1-\mathrm{i}0)+\Omega}
.
\end{eqnarray}
Summation in Eq. (\ref{4b}) extends over the entire Dirac spectrum of electron states $n$ in the field of nucleus. The thermal photon propagator $ D^{\beta}_{\mu\nu}(x_2,x_1) $  can be written in the form \cite{solovyev2015,solovyevarxiv2019}:
\begin{eqnarray}
\label{5b}
D^{\beta}_{\mu\nu}(x_2, x_1) = -  
\frac{g_{\mu\nu}}{\pi r_{12}}\int\limits_{-\infty}^{\infty} n_{\beta}(|\omega_{\beta}|)
\mathrm{sin(|\omega_{\beta}|r_{12})} e^{-\mathrm{i}\omega_{\beta}(t_2-t_1)}d\omega_{\beta}
,
\end{eqnarray}
where $ r_{12}=|\textbf{r}_1-\textbf{r}_2| $ and $ g_{\mu\nu} $ is the metric tensor.

Integration over the time variables in Eq. (\ref{3b}) yields
\begin{eqnarray}
\label{6b}
\hat{S}^{(3)\mathrm{Fig.2}}_{ab}=\frac{(-\mathrm{i}e)^3}{4\pi^3}\int\limits_{-\infty}^{\infty} d\Omega_1 \int\limits_{-\infty}^{\infty} d\Omega_2 \frac{8\eta^3}{((E_a+\Omega_2-\omega)^2+\eta^2)((\Omega_2 - \Omega_1 +\omega_{\beta})^2+\eta^2)((E_{b}+ \Omega_1 -\omega_{\beta})^2+\eta^2)}
\\\nonumber
\times
\sqrt{\frac{2\pi}{\omega}}\int\limits_{-\infty}^{\infty}d\omega_{\beta} n_{\beta}(|\omega_{\beta}|)\sum_{n_2n_1}\frac{\langle n_1|(\textbf{e}^{*}\bm{\alpha})e^{-\mathrm{i}\textbf{k}\textbf{r}}| a \rangle}{E_{n_1}(1-\mathrm{i}0)+\Omega_2}
\frac{\langle bn_2  | \frac{1-\bm{\alpha}_2\bm{\alpha}_3}{r_{23}}\mathrm{sin}(|\omega_{\beta}|r_{23}) |  n_2n_1 \rangle}{E_{n_2}(1-\mathrm{i}0)+\Omega_1}
,
\end{eqnarray}
where integration over variables $ \Omega_{1} $ and $ \Omega_{2} $ in Eq. (\ref{6b}) can be performed with the use of Cauchy theorem. The result is
\begin{eqnarray}
\label{7b}
\int\limits_{-\infty}^{\infty} d\Omega_1 \int\limits_{-\infty}^{\infty} d\Omega_2 \frac{8\eta^3}{((E_a+\Omega_2-\omega)^2+\eta^2)((\Omega_2 - \Omega_1 +\omega_{\beta})^2+\eta^2)((E_{b}+ \Omega_1 -\omega_{\beta})^2+\eta^2)}
\\\nonumber
\times
\frac{1}{E_{n_1}(1-\mathrm{i}0)+\Omega_{2}}\frac{1}{E_{n_2}(1-\mathrm{i}0)+\Omega_{1}}
=\frac{1}{\omega_{n_1b}-2\mathrm{i}\eta}\frac{1}{\omega_{n_2b}+\omega_{\beta}-\mathrm{i}\eta-\mathrm{i}0}\frac{24\pi^2\eta}{(\omega_{ab}-\omega)^2+(3\eta)^2}.
\end{eqnarray}

Following to definition Eq. (\ref{2b}) the correction to one-photon transition rate can be found as
\begin{eqnarray}
\label{8b}
\Delta W_{ab}=\lim\limits_{\eta\rightarrow 0+}\eta\sum_{\textbf{e}}\int 2\mathrm{Re}(\hat{S}^{(1)*}_{ab}\times\hat{S}^{(3)}_{ab})\frac{\omega^2 d\omega d\bm{\nu}}{(2\pi)^3}
.
\end{eqnarray}
Taking in mind that Eq. (\ref{7b}) is multiplied by the expression (\ref{3a}), the integration over photon frequency $\omega$ should be performed
\begin{eqnarray}
\label{9b}
\int\limits_{-\infty}^{+\infty}\frac{2\eta }{(\omega_{ab}-\omega)^2+\eta^2}\frac{24\pi^2\eta }{(\omega_{ab}-\omega)^2+(3\eta)^2}\omega d\omega=\frac{4\pi^3\omega_{ab}}{\eta}
.
\end{eqnarray}
Then Eq. (\ref{8b}) reduces to
\begin{eqnarray}
\label{10b}
\Delta W_{ab}=\mathrm{Re}\left(\omega_{ab}^2\sum_{\textbf{e}}\int \hat{U}^{(1)*}_{ab}\hat{U}^{(3)}_{ab}d\bm{\nu}\right)
,
\end{eqnarray}
where
\begin{eqnarray}
\label{11b}
\hat{U}^{(3)\mathrm{Fig.2}}_{ab}=\frac{(-\mathrm{i}e)^3}{4\pi^3}\sqrt{\frac{2\pi}{\omega_{ab}}}\int\limits_{-\infty}^{\infty}d\omega_{\beta} n_{\beta}(|\omega_{\beta}|)\sum_{n_2n_1}\frac{\langle n_1|(\textbf{e}^{*}\bm{\alpha})e^{-\mathrm{i}\textbf{k}\textbf{r}}| a \rangle}{\omega_{n_1b}-2\mathrm{i}\eta}
\frac{\langle bn_2  | \frac{1-\bm{\alpha}_2\bm{\alpha}_3}{r_{23}}\mathrm{sin}(|\omega_{\beta}|r_{23}) |  n_2n_1 \rangle}{\omega_{n_2b}+\omega_{\beta}-\mathrm{i}\eta}
=
\\\nonumber
=\frac{(-\mathrm{i}e)^3}{4\pi^3}
\sqrt{\frac{2\pi}{\omega_{ab}}}
\int\limits_{0}^{\infty}d\omega_{\beta} n_{\beta}(\omega_{\beta})\sum_{n_2n_1}\frac{\langle n_1|(\textbf{e}^{*}\bm{\alpha})e^{-\mathrm{i}\textbf{k}\textbf{r}}| a \rangle \langle bn_2  | \frac{1-\bm{\alpha}_2\bm{\alpha}_3}{r_{23}}\mathrm{sin}(\omega_{\beta} r_{23}) |  n_2n_1 \rangle}{\omega_{n_1b}-2\mathrm{i}\eta}
\\\nonumber
\times
\left(
\frac{1}{\omega_{n_2b}+\omega_{\beta}-\mathrm{i}\eta}+\frac{1}{\omega_{n_2b}-\omega_{\beta}-\mathrm{i}\eta}\right)
.
\end{eqnarray}

The diagram Fig.~\ref{fig2} and the corresponding amplitude Eq. (\ref{11b}) have reducible part (reference state contribution) when $ n_1=b $. To evaluate it we set $ n_1=b $ in Eq. (\ref{11b}) and consider Taylor expansion in the vicinity of $ \eta=0 $, that gives
\begin{eqnarray}
\label{12b}
\hat{U}^{(3)\mathrm{Fig.2}}_{ab(n_1=b)}=\frac{(-\mathrm{i}e)^3}{4\pi^3}
\sqrt{\frac{2\pi}{\omega_{ab}}}\int\limits_{0}^{\infty}d\omega_{\beta} n_{\beta}(\omega_{\beta})\sum_{n_2}\langle b|(\textbf{e}^{*}\bm{\alpha})e^{-\mathrm{i}\textbf{k}\textbf{r}}| a \rangle \langle bn_2  | \frac{1-\bm{\alpha}_2\bm{\alpha}_3}{r_{23}}\mathrm{sin}(\omega r_{23}) |  n_2b \rangle 
\\\nonumber
\times
\left\lbrace
\frac{\mathrm{i}}{2\eta}\left(
\frac{1}{\omega_{n_2b}+\omega_{\beta}-\mathrm{i}\eta}+\frac{1}{\omega_{n_2b}-\omega_{\beta}-\mathrm{i}\eta}\right)
-\frac{1}{2}
\left(
\frac{1}{(\omega_{n_2b}+\omega_{\beta}-\mathrm{i}\eta)^2}+\frac{1}{(\omega_{n_2b}-\omega_{\beta}-\mathrm{i}\eta)^2}\right)
+O(\eta)
\right\rbrace
.
\end{eqnarray}
The real part (see Eq. (\ref{10b})) of the first term in curly brackets in Eq. (\ref{12b}) multiplied by the expression (\ref{8a}) vanishes with the accounting of pure imaginary factor $1/(-2\mathrm{i}\eta)$.
 Then combination of the reducible and irreducible contributions is
\begin{eqnarray}
\label{13b}
\hat{U}^{(3)\mathrm{Fig.2}}_{ab}=\frac{(-\mathrm{i}e)^3}{4\pi^3}
\sqrt{\frac{2\pi}{\omega_{ab}}}\int\limits_{0}^{\infty}d\omega_{\beta} n_{\beta}(\omega_{\beta})
\left\lbrace
\sum_{n_2,n_1\neq b}\frac{\langle n_1|(\textbf{e}^{*}\bm{\alpha})e^{-\mathrm{i}\textbf{k}\textbf{r}}| a \rangle \langle bn_2  | \frac{1-\bm{\alpha}_2\bm{\alpha}_3}{r_{23}}\mathrm{sin}(\omega_{\beta} r_{23}) |  n_2n_1 \rangle}{\omega_{n_1b}}
\right.
\\\nonumber
\times
\left(
\frac{1}{\omega_{n_2b}+\omega_{\beta}-\mathrm{i}\eta}+\frac{1}{\omega_{n_2b}-\omega_{\beta}-\mathrm{i}\eta}\right)
\\\nonumber
-\frac{1}{2}
\sum_{n_2}\langle b|(\textbf{e}^{*}\bm{\alpha})e^{-\mathrm{i}\textbf{k}\textbf{r}}| a \rangle \langle bn_2  | \frac{1-\bm{\alpha}_2\bm{\alpha}_3}{r_{23}}\mathrm{sin}(\omega_{\beta} r_{23}) |  n_2b \rangle 
\left.
\left(
\frac{1}{(\omega_{n_2b}+\omega_{\beta}-\mathrm{i}\eta)^2}+\frac{1}{(\omega_{n_2b}-\omega_{\beta}-\mathrm{i}\eta)^2}\right)
\right\rbrace
.
\end{eqnarray}

One can note that the thermal corrections are suppressed by the factor of temperature in addition to the $Z\alpha$-expansion. Therefore, applying the nonrelativistic limits to the expression (\ref{13b}) can serve as an adequate approximation for the search for the dominant contribution. Within the dipole approximation $ \textbf{k}\textbf{r} \ll 1 $ we have
\begin{eqnarray}
\label{14b}
\langle n_1|(\textbf{e}^{*}\bm{\alpha})e^{-\mathrm{i}\textbf{k}\textbf{r}}| a \rangle
=\langle n_1|\textbf{e}^{*}\textbf{p}| a \rangle = -\mathrm{i}\omega_{an_1}\langle n_1|\textbf{e}^{*}\textbf{r}| a \rangle
,
\end{eqnarray}
\begin{eqnarray}
\label{15b}
\langle bn_2 |\frac{1-\bm{\alpha}_{2}\bm{\alpha}_{3}}{r_{23}}\mathrm{sin}(\omega_{\beta} r_{23}) | n_2n_1 \rangle
\approx
\omega_{\beta} \langle b | n_2 \rangle \langle n_2 | n_1 \rangle - \omega_{\beta}\langle b |\textbf{p}| n_2 \rangle  \langle n_2 |\textbf{p} |n_1 \rangle + \frac{\omega_{\beta}^3}{3}  \langle b |\textbf{r}| n_2 \rangle\langle n_2 |\textbf{r} |n_1 \rangle
\\\nonumber
=\omega \langle b | n_2 \rangle \langle n_2 | n_1 \rangle  + 
\left(
- \omega\omega_{n_2n_1}\omega_{n_2b}  + \frac{\omega^3}{3} 
\right)
 \langle b |\textbf{r}| n_2 \rangle\langle n_2 |\textbf{r} |n_1 \rangle
, 
\end{eqnarray}
where the relation Eq. (\ref{relation}) was used. Then, substituting Eqs. (\ref{14b}) and (\ref{15b}) into Eq. (\ref{13b}), we find
\begin{eqnarray}
\label{16b}
\hat{U}^{(3)\mathrm{Fig.2}}_{ab}=\frac{(-\mathrm{i}e)^3}{4\pi^3}
\sqrt{\frac{2\pi}{\omega_{ab}}}\int\limits_{0}^{\infty}d\omega_{\beta} n_{\beta}(\omega_{\beta})
\left\lbrace
\sum_{n_2,n_1\neq b}
\frac{-\mathrm{i}\omega_{an_1}\langle n_1|\textbf{e}^{*}\textbf{r}| a \rangle 
}{\omega_{n_1b}}
\right.
\\\nonumber
\times
\left[
\omega_{\beta} \langle b | n_2 \rangle \langle n_2 | n_1 \rangle  
- \omega_{\beta}\langle b |\textbf{p}| n_2 \rangle  \langle n_2 |\textbf{p} |n_1 \rangle 
+ \frac{\omega_{\beta}^3}{3}  \langle b |\textbf{r}| n_2 \rangle\langle n_2 |\textbf{r} |n_1 \rangle
\right]
\left(
\frac{1}{\omega_{n_2b}+\omega_{\beta}-\mathrm{i}\eta}+\frac{1}{\omega_{n_2b}-\omega_{\beta}-\mathrm{i}\eta}\right)
\\\nonumber
-
\frac{1}{2}
\sum_{n_2}(-\mathrm{i}\omega_{ab}\langle b|\textbf{e}^{*}\textbf{r}| a \rangle) 
\left[\omega_{\beta} \langle b | n_2 \rangle \langle n_2 | b \rangle  
- \omega_{\beta}\langle b |\textbf{p}| n_2 \rangle  \langle n_2 |\textbf{p} |b \rangle + \frac{\omega_{\beta}^3}{3}  \langle b |\textbf{r}| n_2 \rangle\langle n_2 |\textbf{r} |b \rangle
\right]
\\\nonumber
\times
\left.
\left(
\frac{1}{(\omega_{n_2b}+\omega_{\beta}-\mathrm{i}\eta)^2}+\frac{1}{(\omega_{n_2b}-\omega_{\beta}-\mathrm{i}\eta)^2}\right)
\right\rbrace
.
\end{eqnarray}

In view of the orthogonality property of wave functions and the nonequlity $ n_1\neq b$ the term $ \omega_{\beta} \langle b | n_2 \rangle \langle n_2 | n_1 \rangle $ turns to zero. In turn, the term $ \omega_{\beta} \langle b | n_2 \rangle \langle n_2 | b \rangle $ leads to infrared divergence for $ n_2=b $:
\begin{eqnarray}
\label{17b}
\hat{U}^{(3)\mathrm{Fig.2}}_{ab(n_2=b)}=-\frac{e^3}{2\pi^{5/2}}\sqrt{\frac{\omega_{ab}}{2}}\int\limits_{0}^{\infty} \frac{d\omega_{\beta}}{\omega_{\beta}} n_{\beta}(\omega_{\beta})
.
\end{eqnarray}
Below we will show that the same divergences occur for the diagrams in Fig. \ref{fig3} and Fig. \ref{fig4}. In particular, from the combination of these three graphs will follow that the singular contribution for the diagram Fig. (\ref{fig4}) is canceled precisely by the corresponding terms in diagrams Fig. (\ref{fig2}) and Fig. (\ref{fig3}). The same has a place to be for the ordinary radiative QED corrections to one-photon transition rates, see \cite{shabaev2000}.
Thus, all the infrared divergences contained in diagrams Figs.~\ref{fig2}-\ref{fig4} are canceled removing the terms $ \omega_{\beta} \langle b | n_2 \rangle \langle n_2 | n_1 \rangle $ and $ \omega_{\beta} \langle b | n_2 \rangle \langle n_2 | b \rangle $ in Eq. (\ref{16b}). Then Eq. (\ref{16b}) reduces to
\begin{eqnarray}
\label{18b}
\hat{U}^{(3)\mathrm{Fig.2}}_{ab}=\frac{(-\mathrm{i}e)^3}{4\pi^3}
\sqrt{\frac{2\pi}{\omega_{ab}}}\int\limits_{0}^{\infty}d\omega_{\beta} n_{\beta}(\omega_{\beta})
\left\lbrace
\sum_{n_2,n_1\neq b}
\frac{-\mathrm{i}\omega_{an_1}\langle n_1|\textbf{e}^{*}\textbf{r}| a \rangle 
}{\omega_{n_1b}}
\right.
\\\nonumber
\times
\left[
- \omega_{\beta}\langle b |\textbf{p}| n_2 \rangle  \langle n_2 |\textbf{p} |n_1 \rangle + \frac{\omega_{\beta}^3}{3}  \langle b |\textbf{r}| n_2 \rangle\langle n_2 |\textbf{r} |n_1 \rangle
\right]
\left(
\frac{1}{\omega_{n_2b}+\omega_{\beta}-\mathrm{i}\eta}+\frac{1}{\omega_{n_2b}-\omega_{\beta}-\mathrm{i}\eta}\right)
\\\nonumber
-
\frac{1}{2}
\sum_{n_2}(-\mathrm{i}\omega_{ab}\langle b|\textbf{e}^{*}\textbf{r}| a \rangle) 
\left[ 
- \omega_{\beta}\langle b |\textbf{p}| n_2 \rangle  \langle n_2 |\textbf{p} |b \rangle + \frac{\omega_{\beta}^3}{3}  \langle b |\textbf{r}| n_2 \rangle\langle n_2 |\textbf{r} |b \rangle
\right]
\left.
\left(
\frac{1}{(\omega_{n_2b}+\omega_{\beta}-\mathrm{i}\eta)^2}+\frac{1}{(\omega_{n_2b}-\omega_{\beta}-\mathrm{i}\eta)^2}\right)
\right\rbrace
.
\end{eqnarray}

The expression (\ref{18b}) can be rewritten in another way:
\begin{eqnarray}
\label{18bSTARK}
\hat{U}^{(3)\mathrm{Fig.2}}_{ab}=\frac{(-\mathrm{i}e)^3}{4\pi^3}
\sqrt{\frac{2\pi}{\omega_{ab}}}\int\limits_{0}^{\infty}d\omega_{\beta} n_{\beta}(\omega_{\beta})
\left\lbrace
\sum_{n_2,n_1\neq b}
\frac{-\mathrm{i}\omega_{an_1}\langle n_1|\textbf{e}^{*}\textbf{r}| a \rangle 
}{\omega_{n_1b}}
\right.
\\\nonumber
\times
\left[
- \omega_{\beta}\langle b |\textbf{p}| n_2 \rangle  \langle n_2 |\textbf{p} |n_1 \rangle + \frac{\omega_{\beta}^3}{3}  \langle b |\textbf{r}| n_2 \rangle\langle n_2 |\textbf{r} |n_1 \rangle
\right]
\left(
\frac{1}{\omega_{n_2b}+\omega_{\beta}-\mathrm{i}\eta}+\frac{1}{\omega_{n_2b}-\omega_{\beta}-\mathrm{i}\eta}\right)
\\\nonumber
-
\frac{1}{2}
\sum_{n_2}(-\mathrm{i}\omega_{ab}\langle b|\textbf{e}^{*}\textbf{r}| a \rangle) 
\frac{\partial}{\partial E_{b}}
\left[ 
- \omega_{\beta}\langle b |\textbf{p}| n_2 \rangle  \langle n_2 |\textbf{p} |b \rangle + \frac{\omega_{\beta}^3}{3}  \langle b |\textbf{r}| n_2 \rangle\langle n_2 |\textbf{r} |b \rangle
\right]
\left.
\left(
\frac{1}{\omega_{n_2b}+\omega_{\beta}-\mathrm{i}\eta}+\frac{1}{\omega_{n_2b}-\omega_{\beta}-\mathrm{i}\eta}\right)
\right\rbrace
.
\end{eqnarray}
Then, applying relation (\ref{relation}), we obtain the final expression within the nonrelativistic limit for the diagram in Fig.~\ref{fig2}:
\begin{eqnarray}
\label{19b}
\hat{U}^{(3)\mathrm{Fig.2}}_{ab}=\frac{(-\mathrm{i}e)^3}{4\pi^3}
\sqrt{\frac{2\pi}{\omega_{ab}}}\int\limits_{0}^{\infty}d\omega_{\beta} n_{\beta}(\omega_{\beta})
\left\lbrace
\sum_{n_2,n_1\neq b}\frac{-\mathrm{i}\omega_{an_1}\langle n_1|\textbf{e}^{*}\textbf{r}| a \rangle 
\langle b |\textbf{r}| n_2 \rangle\langle n_2 |\textbf{r} |n_1 \rangle
}{\omega_{n_1b}}
\right.
\left(
- \omega_\beta\omega_{n_2n_1}\omega_{n_2b}  + \frac{\omega_\beta^3}{3} 
\right)
\\\nonumber
\times
\left(
\frac{1}{\omega_{n_2b}+\omega_{\beta}-\mathrm{i}\eta}+\frac{1}{\omega_{n_2b}-\omega_{\beta}-\mathrm{i}\eta}\right)
-
\left.
\frac{\mathrm{i}\omega_{ab}\langle b|\textbf{e}^{*}\textbf{r}| a \rangle}{2}
\frac{\partial}{\partial   E_{b}}
\sum_{n_2} \langle b |\textbf{r}| n_2 \rangle\langle n_2 |\textbf{r} |b\rangle
\left(\frac{4\omega_{\beta}^3}{3} 
\frac{\omega_{n_2b}}{\omega_{n_2b}^2-\omega_{\beta}^2}\right)
\right\rbrace
.
\end{eqnarray}

Evaluation of the diagram in Fig.~\ref{fig3} repeats the procedure above, see Eqs. (\ref{7b})-(\ref{19b}), with the S-matrix element
\begin{eqnarray}
\label{20b}
\hat{S}^{(3)\mathrm{Fig.3}}_{ab}=(-\mathrm{i}e)^3\int dx_{3}dx_{2}dx_{1}\overline{\psi}_{b}(x_3)\gamma_{\mu_3}A^{*}_{\mu_3}(x_3)S(x_3,x_2)e^{-\eta |t_3|}D^{\beta}_{\mu_2\mu_1}(x_2,x_1)e^{-\eta |t_2|}\gamma_{\mu_2}S(x_2,x_1)\gamma_{\mu_1}e^{-\eta |t_1|}\psi_{a}(x_1)
.
\end{eqnarray}
Then the transition amplitude can be written as
\begin{eqnarray}
\label{21b}
\hat{U}^{(3)\mathrm{Fig.3}}_{ab}=\frac{(-\mathrm{i}e)^3}{4\pi^3}
\sqrt{\frac{2\pi}{\omega_{ab}}}\int\limits_{0}^{\infty}d\omega_{\beta} n_{\beta}(\omega_{\beta})
\left\lbrace
\sum_{n_1,n_2\neq a}\frac{-\mathrm{i}\omega_{n_2b}\langle b|\textbf{e}^{*}\textbf{r}| n_2 \rangle 
\langle n_2 |\textbf{r}| n_1 \rangle\langle n_1 |\textbf{r} |a \rangle
}{\omega_{n_2a}}
\right.
\left(
- \omega_\beta\omega_{n_2n_1}\omega_{an_1}  + \frac{\omega_\beta^3}{3} 
\right)
\\\nonumber
\times
\left(
\frac{1}{\omega_{n_1a}+\omega_{\beta}-\mathrm{i}\eta}+\frac{1}{\omega_{n_1a}-\omega_{\beta}-\mathrm{i}\eta}\right)
-
\left.
\frac{\mathrm{i}\omega_{ab}\langle b|\textbf{e}^{*}\textbf{r}| a \rangle}{2}
\frac{\partial}{\partial   E_{a}}
\sum_{n_1} \langle a |\textbf{r}| n_1 \rangle\langle n_1 |\textbf{r} |a\rangle
\left(\frac{4\omega_{\beta}^3}{3} 
\frac{\omega_{n_1a}}{\omega_{n_1a}^2-\omega_{\beta}^2}
\right)
\right\rbrace
,
\end{eqnarray}
where the same infrared divergence as in Eq. (\ref{17b}) has arises with $n_1=a$. 

Now we can consider the last diagram given by Fig. \ref{fig4}. The corresponding S-matrix element is
\begin{eqnarray}
\label{22b}
\hat{S}^{(3)\mathrm{Fig.4}}_{ab}=(-\mathrm{i}e)^3\int dx_{3}dx_{2}dx_{1}\overline{\psi}_{b}(x_3)S(x_3,x_2)e^{-\eta |t_3|}\gamma_{\mu_3}D^{\beta}_{\mu_3\mu_1}(x_3,x_1)\gamma_{\mu_2}A^{*}_{\mu_2}(x_2)\gamma_{\mu_1}e^{-\eta |t_2|}S(x_2,x_1)e^{-\eta |t_1|}\psi_{a}(x_1),
\end{eqnarray}
Performing integration over time variables and frequencies in the electron and photon propagators, we arrive at the following expression for the transition amplitude:
\begin{eqnarray}
\label{23b}
\hat{U}^{(3)\mathrm{Fig.4}}_{ab}=\frac{(-\mathrm{i}e)^3}{4\pi^3}
\sqrt{\frac{2\pi}{\omega_{ab}}}
\left\lbrace\sum_{n_2n_1}(-\mathrm{i})\omega_{n_1n_2} \langle n_2|\textbf{e}^{*}\textbf{r}| n_1 \rangle
\int\limits_{0}^{\infty}d\omega_{\beta} n_{\beta}(\omega_{\beta})
\left[\omega_{\beta}
\langle b | n_2 \rangle \langle n_1 | a\rangle
\right.
\right.
\\\nonumber
\left.
\left.
+
\left( -\omega_{\beta}\omega_{bn_2}\omega_{an_1}+\frac{\omega_{\beta}^3}{3} \right)
\langle b |\textbf{r}|n_2 \rangle \langle n_1 |\textbf{r}|a\rangle 
\right]
\right.
\\\nonumber
\times
\left.
\left(\frac{1}{(\omega_{n_2b}+\omega_{\beta}-\mathrm{i}\eta)(\omega_{n_1a}+\omega_{\beta}-\mathrm{i}\eta)} + \frac{1}{(\omega_{n_2b}-\omega_{\beta}-\mathrm{i}\eta)(\omega_{n_1a}-\omega_{\beta}-\mathrm{i}\eta)} \right)\right\rbrace
.
\end{eqnarray}
The infrared divergence in Eq. (\ref{23b}) appears from the term $ \omega \langle b | n_2 \rangle \langle n_1 | a\rangle $ with $ n_1=a $ and $ n_2=b $. Then the divergent part of the amplitude Fig. \ref{fig4} congregates to
\begin{eqnarray}
\label{24b}
\hat{U}^{(3)\mathrm{Fig.4}}_{ab(n_1=a,n_2=b)}=\frac{e^3}{\pi^{5/2}}\sqrt{\frac{\omega_{ab}}{2}}\int\limits_{0}^{\infty} \frac{d\omega_{\beta} n_{\beta}(\omega_{\beta})}{\omega_{\beta}}.
\end{eqnarray}
The result (\ref{24b}) has the opposite sign and is twice larger than in Eq. (\ref{17b}) canceling the aggregated contribution Eq. (\ref{17b}) for the diagrams Fig.~\ref{fig2}, \ref{fig3}. 
Then, Eq. (\ref{23b}) can be written in the form:
\begin{eqnarray}
\label{25b}
\hat{U}^{(3)\mathrm{Fig.4}}_{ab}=\frac{(-\mathrm{i}e)^3}{4\pi^3}
\sqrt{\frac{2\pi}{\omega_{ab}}}
\left\lbrace\sum_{n_2n_1}(-\mathrm{i})\omega_{n_1n_2} 
\langle n_2|\textbf{e}^{*}\textbf{r}| n_1 \rangle
\langle b |\textbf{r}|n_2 \rangle 
\langle n_1 |\textbf{r}|a\rangle
\int\limits_{0}^{\infty}d\omega_{\beta} n_{\beta}(\omega_{\beta})
\left(
 -\omega_{\beta}\omega_{bn_2}\omega_{an_1}+\frac{\omega_{\beta}^3}{3}
\right) 
\right.
\\\nonumber
\times
\left.
\left[\frac{1}{(\omega_{n_2b}+\omega_{\beta}-\mathrm{i}\eta)(\omega_{n_1a}+\omega_{\beta}-\mathrm{i}\eta)} + \frac{1}{(\omega_{n_2b}-\omega_{\beta}-\mathrm{i}\eta)(\omega_{n_1a}-\omega_{\beta}-\mathrm{i}\eta)} \right]\right\rbrace
.
\end{eqnarray}

Finally, the thermal radiative corrections corresponding to the diagrams in Fig.~\ref{fig2}-\ref{fig4}, after the integration over photon directions and summation over polarizations, are given by
\begin{gather}
\label{F1}
\Delta W^{\mathrm{Fig.2}}_{ab}=-\frac{4e^4}{3\pi}
\frac{1}{2l_{a}+1}
\sum_{m_{a}m_{b}}
\sum\limits_{n_2n_1}
\left\lbrace\sum_{n_2,n_1\neq b }\frac{\omega_{ab}^2\omega_{an_1}\langle a|\textbf{r}| b \rangle \langle n_1|\textbf{r}| a \rangle
\langle b |\textbf{r}| n_2 \rangle\langle n_2 |\textbf{r} |n_1 \rangle 
}{\omega_{n_1b}}\mathcal{P}\int\limits_{0}^{\infty}d\omega_{\beta} n_{\beta}(\omega_{\beta}) 
\right.
\\\nonumber
\times
\left(
-\omega_{\beta}\omega_{n_2n_1}\omega_{n_2b}+\frac{\omega^3_{\beta}}{3}
\right) 
\left(
\frac{2\omega_{n_2b}}{\omega_{n_2b}^2-\omega_{\beta}^2}
\right)
-\frac{\omega_{ab}^3|\langle a|\textbf{r}| b \rangle |^2 |\langle b |\textbf{r}| n_2 \rangle  |^2
}{2}
\frac{\partial}{\partial E_{b}}\mathcal{P}\int\limits_{0}^{\infty}d\omega_{\beta} n_{\beta}(\omega_{\beta})\sum_{n_2}  
\left.
\left(
\frac{4\omega_{\beta}^3}{3}
\frac{\omega_{n_2b}}{\omega_{n_2b}^2-\omega_{\beta}^2}
\right)\right\rbrace
\end{gather}
,
\begin{gather}
\label{F2}
\Delta W^{\mathrm{Fig.3}}_{ab}=-\frac{4e^4}{3\pi}
\frac{1}{2l_{a}+1}
\sum_{m_{a}m_{b}}
\sum\limits_{n_2n_1}
\left\lbrace\sum_{n_1,n_2\neq a}\frac{\omega_{ab}^2\omega_{n_2b}
\langle a|\textbf{r}| b \rangle \langle b|\textbf{r}| n_2 \rangle
\langle n_2 |\textbf{r} |n_1 \rangle \langle n_1 |\textbf{r}| a \rangle
}{\omega_{n_2a}}\mathcal{P}\int\limits_{0}^{\infty}d\omega_{\beta} n_{\beta}(\omega_{\beta}) 
\right.
\\\nonumber
\times
\left(
-\omega_{\beta}\omega_{n_2n_1}\omega_{an_1}+\frac{\omega^3_{\beta}}{3}
\right) 
\left(
\frac{2\omega_{n_1a}}{\omega_{n_1a}^2-\omega_{\beta}^2}
\right)
-\frac{\omega_{ab}^3|
\langle a|\textbf{r}| b \rangle |^2|\langle a |\textbf{r}| n_1 \rangle  |^2}{2}
\frac{\partial }{\partial E_{a}}\mathcal{P}\int\limits_{0}^{\infty}d\omega_{\beta} n_{\beta}(\omega_{\beta})\sum_{n_1} 
\left.
\left(
\frac{4\omega_{\beta}^3}{3} 
\frac{\omega_{n_1a}}{\omega_{n_1a}^2-\omega_{\beta}^2}
\right)
\right\rbrace
\end{gather}
,
\begin{gather}
\label{F3}
\Delta W^{\mathrm{Fig.4}}_{ab}=-\frac{4e^4}{3\pi}\frac{1}{2l_{a}+1}\sum_{m_{a}m_{b}}
\sum\limits_{n_2n_1}\omega_{ab}^2\omega_{n_1n_2}
\langle a|\textbf{r}| b \rangle \langle n_2|\textbf{r}| n_1 \rangle
\langle b |\textbf{r}|n_2 \rangle \langle n_1 |\textbf{r}|a\rangle 
\mathcal{P}\int\limits_{0}^{\infty}d\omega_{\beta} n_{\beta}(\omega_{\beta})
\\\nonumber
\times
\left(
 -\omega_{\beta}\omega_{n_2b}\omega_{n_1a}+\frac{\omega_{\beta}^3}{3}
\right) 
\frac{2(\omega_{\beta}^2+\omega_{n_2b}\omega_{n_1a})}{(\omega_{n_2b}^2-\omega_{\beta}^2)(\omega_{n_1a}^2-\omega_{\beta}^2)}
.
\end{gather}
Here the principal value of integral, $\mathcal{P}$, has appeared as a result of taking the real part in Eq. (\ref{10b}) and application of Sokhotski-Plemelj theorem in the limit $\eta\rightarrow 0$.

\begin{figure}
\includegraphics[scale=0.7]{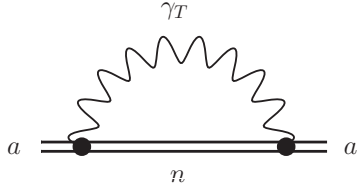}
\caption{The Feynman graph corresponding to the thermal one-loop electron self-energy. The wavy line with index $ \gamma_{T} $ denotes the thermal photon propagator. The real part of this graph corresponds to the thermal Stark shift, while the imaginary part represents the induced one-photon width $ \Gamma^{\mathrm{BBR}}_{a} $ of the atomic energy level $ a $.}
\label{fig5}
\end{figure}

As a next step in evaluating thermal radiative corrections of the type as in Fig.~\ref{fig2}-\ref{fig4} the contribution of thermal one-loop diagram to the energy levels $a$ and $b$ should be considered, see Fig.~\ref{fig5}.
Recently this graph was evaluated in \cite{solovyev2015}, where the real part representing the Stark shift was found in complete agreement with \cite{farley}:
\begin{eqnarray}
\label{28b}
\Delta E_{a}^{\beta}=\frac{1}{2l_{a}+1}\sum_{m_{a}m_{b}}\frac{4e^2}{3\pi}\sum\limits_{n}|\langle a| \textbf{r}|n\rangle |^2\mathcal{P}\int\limits_{0}^{\infty}d\omega_{\beta} \omega_{\beta}^3 n_{\beta}(\omega_{\beta})\frac{\omega_{an}}{\omega_{an}^2-\omega_{\beta}^2}
.
\end{eqnarray}
Numerical values of the BBR-induced Stark shift at different temperatures are collected in Table~\ref{tab1} for the $1s$, $2s$ and $2p$ states in hydrogen atom.
\begin{table}[hbtp]
\caption{Blackbody-radiation-induced dynamic Stark shifts (Hz) of energy levels of hydrogen at different temperatures (Kelvin). Fine structure and the Lamb shift were neglected in performing the calculation. The asterisks corresponds to the values presented in \cite{farley}.}
\begin{tabular}{c c c c}
\hline
  $T$   & $ 1s $ & $ 2s $ & $ 2p $\\
\hline
300   & -3.8754[-2] & -1.0434     & -1.5181   \\
      & -4.128[-2]$ ^* $  & -1.077$ ^* $&  -1. 535$ ^* $  \\
1000  & -4.7893     & -1.3229[2]  & -1.9438[2]\\
3000  & -3.9148[2]  & -1.4111[4]  & -2.1033[4]\\
5000  & -3.0797[2]  & -7.6499[4]  & -1.0774[5]\\
8000  & -2.1370[4]  & -1.2917[5]  & -1.2786[5]\\
10000 & -5.5601[4]  & -3.0019[4]  &  9.3418[4]\\
50000 & -2.5583[-6] &  4.5616[7]  &  6.0049[7]\\
\hline
\end{tabular}
\label{tab1}
\end{table}
The real part of the diagram in Fig.~\ref{fig5} for the states $a$ and $b$ contributes to the transition rate, see \cite{shabaev2000}, as follows
\begin{eqnarray}
\label{29b}
\Delta W^{\mathrm{Stark}}_{ab}=\frac{1}{2l_{a}+1}\sum_{m_{a}m_{b}}\frac{4e^2}{3}(\omega^3_{ab}-\tilde{\omega}^3_{ab})|\langle b |\textbf{r}|a\rangle|^2
,
\end{eqnarray}
where $ \tilde{\omega}_{ab}=E_{a}+\Delta E^{\beta}_{a} - E_{b}- \Delta E^{\beta}_{b} $. 

Finally, the total thermal QED correction, arising from the graphs in Figs.~\ref{fig2}-\ref{fig5}, to the one-photon transition rate is given by
\begin{eqnarray}
\label{30b}
\Delta W^{\mathrm{total}}_{ab}=
\Delta W^{\mathrm{Fig.2}}_{ab}+\Delta W^{\mathrm{Fig.3}}_{ab}+\Delta W^{\mathrm{Fig.4}}_{ab}+\Delta W^{\mathrm{Stark}}_{ab}
.
\end{eqnarray}
Here $\Delta W^{\mathrm{Fig.2}}_{ab}$, $\Delta W^{\mathrm{Fig.3}}_{ab}$, $\Delta W^{\mathrm{Fig.4}}_{ab}$ and $\Delta W^{\mathrm{Stark}}_{ab}$ are defined by Eqs. (\ref{F1})-(\ref{F3}) and (\ref{29b}), respectively. These corrections were obtained in the nonrelativistic limit as the contributions of leading order in $ \alpha $. Numerical values of $\Delta W^{\mathrm{Fig.2}}_{ab}$, $\Delta W^{\mathrm{Fig.3}}_{ab}$, $\Delta W^{\mathrm{Fig.4}}_{ab}$ and $\Delta W^{\mathrm{Stark}}_{ab}$ for the $2p\rightarrow 1s +\gamma(\mathrm{E1})$ transition in the hydrogen atom at different temperatures are collected in Table~\ref{tab2}.
\begin{table}[hbtp]
\caption{Corrections to one-photon transition rates corresponding to Figs. \ref{fig2}-\ref{fig4}, induced transition rates  $ W^{\mathrm{ind}}_{2p1s}  $ and BBR-induced width $ \Gamma_{2p}^{\mathrm{BBR}} $ in s$^{-1}  $ at different temperatures (Kelvin).}
\begin{tabular}{c c c c c c c c} 
\hline
  $T$  & $ \Delta W^{\mathrm{Fig.2}}_{2p1s} $ & $ \Delta W^{\mathrm{Fig.3}}_{2p1s} $ & $ \Delta W^{\mathrm{Fig.4}}_{2p1s} $ & $ \Delta W^{\mathrm{Stark}}_{2p1s} $ & $ \Delta W^{\mathrm{total}}_{2p1s} $ & $ W^{\mathrm{ind}}_{2p1s} $ & $ \Gamma_{2p}^{\mathrm{BBR}} $\\
  \hline
300   & -6.3948[-5] & 1.5111[-3] & 9.7273[-4]   & 1.1547[-6]  & 2.4210[-3] & 2.3598[-163]& 4.7430[-6]\\
1000  & -7.1011[-4] & 1.7011[-2] & 1.0883[-2]   & 1.4450[-4]  & 2.7328[-2] & 2.3435[-43] & 3.2943[-2]\\
3000  & -6.3567[-3] & 1.7575[-1] & 1.0602[-1]   & 1.5731[-2]  & 2.9114[-1] & 4.5156[-9]  & 7.5856[4]\\
5000  & -1.7445[-2] & 4.1682[-1] & 3.1890[-1]   & 7.9768[-2]  & 7.9804[-1] & 3.2486[-9]  & 1.5306[6]\\
8000  & -4.0087[-2] & 6.6053[-1] & 7.83286[-1]  &  8.1160[-2] & 1.4849     & 2.3373[2]   & 9.2948[6]\\    
10000 & -6.3120[-2] & 8.4717[-1] & 1.1223       & -1.1357[-1] & 3.3473[-1] & 4.5126[3]   & 1.7960[7]\\
50000 &  -5.1929    &  2.8715[1] & -3.7875      &  -4.7716[1] &-2.79814[1] & 6.4759[7]   &    
4.4084[8]\\
\hline
\end{tabular}
\label{tab2}
\end{table}

\section{Conclusions and discussion}
\label{theend}

In this work, the thermal one-loop self-energy corrections to the one-photon transition rate were obtained within the framework of QED theory. The partial transition probabilities and total depopulation rates induced by the BBR field, see Eqs. (\ref{10a}) and (\ref{11a}), at different temperatures were evaluated also. The numerical values are given in Table~\ref{tab2}. For the summation over entire spectrum in Eqs. (\ref{F1})-(\ref{F3}) the B-spline method was employed \cite{dkb}. Spectrum of virtual states was checked on calculation of Stark shifts and depopulation rates, see Tables~\ref{tab1}, \ref{tab2}. As an additional verification, the Thomas-Reiche-Kuhn sum rule was also checked by this method. Results of calculations for BBR-induced Stark shifts at $ T=300 $ K are in excellent agreement with the our previous results \cite{solovyev2015}. However at high temperatures the present calculations are more accurate due to the improved numerical integration over $ \omega_{\beta} $ in Eq. (\ref{28b}). 

As an example we focused on the numerical evaluation of Eq. (\ref{30b}) for the Ly$_{\alpha}$ $2p\rightarrow 1s +\gamma(\mathrm{E1})$ transition in hydrogen atom, see Table~\ref{tab2}. To demonstrate the role of thermal QED corrections it is useful to compare them with the 'ordinary' QED corrections. The radiative QED correction of lowest order to Ly$ _{\alpha} $ transition was found in \cite{saperstein} as
\begin{eqnarray}
\label{ordinaryQED}
\Delta W^{\mathrm{QED}}_{2p1s}=W_{2p1s}\frac{\alpha}{\pi}(\alpha Z)^2
\left[
\left(
\frac{8}{3}\mathrm{ln}\frac{4}{3}-\frac{61}{18}
\right)
\mathrm{ln}(\alpha Z)^{-2}
+6.57603
\right]
\approx -1490\;\mathrm{s}^{-1}
,
\end{eqnarray}
where $ W_{2p1s}=6.268\times 10^{8} $ s$ ^{-1} $ is the spontaneous one-photon decay rate of lowest order, see Eq. (\ref{9aa}). As can be seen from Table \ref{tab2}, the thermal radiative SE corrections reach the same order starting from $T \sim 50000$ K. Then it can be expected that these corrections can play a role in various astrophysical processes (the case of high temperatures). The results are applicable up to the hydrogen ionization temperature $T_{\mathrm{ion}}\sim 157000$ K, when corrections Eqs. (\ref{F1})-(\ref{F3}) and (\ref{29b}) may be even more important than the radiative QED effects Eq. (\ref{ordinaryQED}).

Actually, it is more correct ещ analyze these thermal contributions with respect to the BBR-induced transition rates.  In particular, from Eqs. (\ref{F1})-(\ref{F3}) follows that the parametric estimation of the leading order in $\alpha$ for the thermal contributions $\Delta W^{\mathrm{Fig.2}}_{ab}$, $\Delta W^{\mathrm{Fig.3}}_{ab}$ and $\Delta W^{\mathrm{Fig.4}}_{ab}$ is the same as a simplest correction arising via the Stark shift, Eq. (\ref{29b}). Then the magnitudes of $ \Delta W^{\mathrm{Fig.2}}_{ab}$, $\Delta W^{\mathrm{Fig.3}}_{ab}$ and $\Delta W^{\mathrm{Fig.4}}_{ab}$ at the room temperature are about hundred times larger than the induced transition rate, $ W^{\mathrm{ind}}_{2p1s} $. They become comparable at the temperatures about $T=6000-8000$ K. Beginning from $8000$ K the thermal radiative corrections are less than the BBR-induced transition rate $  W^{\mathrm{ind}}_{2p1s} $. From Table~\ref{tab2} follows also that the BBR-induced line broadening $ \Gamma_{2p}^{\mathrm{BBR}} $ is the same order of magnitude as the corrections in Eq. (\ref{30b}) at the temperature $T=1000$ K and exceeds $ \Delta W^{\mathrm{total}}_{2p1s} $ at higher temperatures. 

Concluding, a new type of radiative and temperature dependent corrections to the decay rates $\Delta W^{\mathrm{Fig.2}}_{ab}$, $\Delta W^{\mathrm{Fig.3}}_{ab}$, $\Delta W^{\mathrm{Fig.4}}_{ab}$ and $\Delta W^{\mathrm{Stark}}_{ab}$, see Eqs. (\ref{F1})-(\ref{F3}) and (\ref{29b}), were introduced. These corrections may be important in laboratory and astrophysical problems dealing with the BBR-induced effects. Knowledge of the lifetimes and transition probabilities plays an important role in modeling various astrophysical processes, such as the theoretical description of cosmic microwave background (CMB) \cite{chluba} for example. The role of the effects considered in this paper can be emphasized by the problem of a detailed description of the history of the early universe, since radiation temperatures at the early stage of CMB formation correspond to several thousand Kelvin.

\section*{Acknowledgements}
The work of T. Z. was supported by the Foundation for the Advancement of Theoretical Physics and Mathematics ``BASIS``. The work was also supported by RFBR grant Nr. 20-02-00111.


\begin{thebibliography}{100}

\bibitem{drake} G. W. F. Drake, Phys. Rev. A {\bf 9}, 2799 (1974).

\bibitem{sucher} J. Sucher, Phys. Rev. {\bf 107}, 1448 (1957).

\bibitem{saperstein} J. Sapirstein, K. Pachucki, and K. T. Cheng, Phys. Rev. A {\bf 69}, 022113 (2004).

\bibitem{volotka} A. V. Volotka, D. A. Glazov, G. Plunien, V. M. Shabaev, I. I. Tupitsyn, Eur. Phys. J. D {\bf 38}, 293-298 (2006).

\bibitem{st1} T. V. Back, H. S. Margolis, P. K. Oxley, J. D. Silver, and E. G. Myers, Hyperfine Int. {\bf 114}, 203 (1998).

\bibitem{st2} D. P. Moehs and D. A. Church, Phys. Rev. A {\bf 58}, 1111 (1998).

\bibitem{st3} E. Tr\"abert, G. Gwinner, A. Wolf, X. Tordoir, and A. G. Calamai, Phys. Lett. A {\bf 264}, 311 (1999).

\bibitem{st4} E. Tr\"abert, P. Beiersdorfer, S. B. Utter, G. V. Brown, H. Chen, C. L. Harris, P. A. Neill, D. W. Savin, and A. J. Smith, Astrophys. J. {\bf 541}, 506 (2000).

\bibitem{st5} E. Tr\"abert, P. Beiersdorfer, G. V. Brown, H. Chen, E. H. Pinnington, and D. B. Thorn, Phys. Rev. A {\bf 64}, 034501 (2001).

\bibitem{st6} E. Tr\"abert, P. Beiersdorfer, G. Gwinner, E. H. Pinnington, and A. Wolf, Phys. Rev. A {\bf 66}, 052507 (2002).

\bibitem{st7} J. R. Crespo Lopez-Urrutia, A. N. Artemyev, J. Braun, G. Brenner, H. Bruhns, I. N. Draganic, A. J.Gonzalez Martinez, A. Lapierre, V. Mironov, J. Scofield, R. Soria Orts, H. Tawara, M. Trinczek, I. I.Tupitsyn, and J. Ullrich, Nucl. Instr. Meth. Phys. Res. B {\bf 235}, 85 (2005).

\bibitem{tupvol2} A. Lapierre, U. D. Jentschura, J. R. Crespo López-Urrutia, J. Braun, G. Brenner, H. Bruhns, D. Fischer, A. J. Gonzalez Martínez, Z. Harman, W. R. Johnson, C. H. Keitel, V. Mironov, C. J. Osborne, G. Sikler, R. Soria Orts, V. Shabaev, H. Tawara, I. I. Tupitsyn, J. Ullrich, and A. Volotka, Phys. Rev. Lett. {\bf 95}, 183001 (2005).

\bibitem{rec1} Z. Fried and A. O. Martin, Nuovo Cim. 23, {\bf 574} (1963).

\bibitem{rec2} G. S. Adkins and J. Sapirstein, Phys. Rev. A {\bf 78}, 062503
(2008).

\bibitem{rec3} S. G. Karshenboim, Phys. Rev. A {\bf 56}, 4311 (1997).

\bibitem{tup} I. I. Tupitsyn, A. V. Volotka, D. A. Glazov, V. M. Shabaev, G. Plunien, J. R. Crespo Lopez-Urrutia, A. Lapierre, J. Ullrich, Phys. Rev. A {\bf 72}, 062503 (2005).

\bibitem{drakerel} G. W. F. Drake, Phys. Rev. A {\bf 5}, (1979).

\bibitem{source} H. F. Beyer and V. P. Shevelko, {\it Introduction to the Physics
of Highly Charged Ions}, Inst. Physics, Bristol, Philadelphia, (2003).

\bibitem{saper2} . J. Sapirstein, K. Pachucki, A. Veitia, and K. T. Cheng,
Phys. Rev. A {\bf 67}, 052110 (2003).

\bibitem{shab0} . V. M. Shabaev, I. I. Tupitsyn, K. Pachucki, G. Plunien,
and V. A. Yerokhin, Phys. Rev. A 72, 062105 (2005).

\bibitem{tupvol1} I. I. Tupitsyn, A. V. Volotka, D. A. Glazov, V. M. Shabaev, G. Plunien, J. R. Crespo Lopez-Urrutia, A. Lapierre, and J. Ullrich
Phys. Rev. A 72, 062503


\bibitem{zalialiutdinov} T. Zalialiutdinov, D. Solovyev, L. Labzowsky, and G. Plunien
Phys. Rev. A {\bf 89} 052502 (2014).

\bibitem{jent} B. J. Wundt, U. D. Jentschura, Phys. Rev. A {\bf 80}, 022505 (2009).

\bibitem{riehle} F. Riehle, {\it Frequency Standards: Basics and Applications}, WILEY-VCH Verlag GmbH \& Co. KGaA, Weinheim (2004).

\bibitem{labbook} L. N. Labzowsky, G. L. Klimchitskaya an Yu. Yu. Dmitriev, {\it Relativistic effects in the spectra of atomic systems}, IOP Publishing, Bristol and Philadelphia (1993).

\bibitem{fradkin} E. S. Fradkin, D. M. Guitman, S. M. Shvartsman, {\it Quantum Electrodynamics
with Unstable Vacuum},  Springer-Verlag Berlin Heidelberg (1991).

\bibitem{kaplan} S.A. Kaplan, S.B. Pikelner, {\it The Interstellar Medium}, Harvard University Press, Cambridge, (1970).

\bibitem{sol-ext} D. Solovyev, V. Sharipov, L. Labzowsky and G. Plunien, J. Phys. B: At. Mol. Opt. Phys. {\bf 43}, 074005 (2010)

\bibitem{safronova1} K. W. Martin, B. Stuhl, J. Eugenio, M. S. Safronova, G. Phelps, J. H. Burke, N. D. Lemke,
Phys. Rev. A {\bf 100}, 023417 (2019).

\bibitem{middelmann} T. Middelmann, S. Falke, C. Lisdat, U. Sterr, Phys. Rev. Lett. {\bf 109}, 263004 (2012).

\bibitem{safronova2} M. S. Safronova, S. G. Porsev, U. I. Safronova, M. G. Kozlov, and C. W. Clark
Phys. Rev. A {\bf 87}, 012509 (2013).

\bibitem{beterov} I. I. Beterov, D. B. Tretyakov, I. I. Ryabtsev, V. M. Entin, A. Ekers, N. N. Bezuglov, New J. Phys. {\bf 11}, 013052 (2009).

\bibitem{ovsiannikov} V. D. Ovsiannikov, I. L. Glukhov, E. A. Nekipelov, J. Phys. B: At. Mol. Opt. Phys. {\bf 44}, 195010 (2011).

\bibitem{chluba} J. Chluba, R. A. Sunyaev, A\& A {\bf 446}, 1, 39-42 (2006).

\bibitem{hirata} C. M. Hirata, Phys. Rev. D {\bf 78}, 023001 (2008).

\bibitem{farley} J. W. Farley, W. H. Wing, Phys. Rev. A {\bf 23}, 2397 (1981).


\bibitem{solovyev2015} D. Solovyev, L. Labzowsky, and G. Plunien, Phys. Rev. A {\bf 92}, 022508 (2015).

\bibitem{zalialiutdinov2018} T. Zalialiutdinov, D. Solovyev, L. Labzowsky, J. Phys. B: At. Mol. Opt. Phys. {\bf 51}, 015003 (2018).

\bibitem{zalialitdinov2019} T. Zalialiutdinov, D. Solovyev, L. Labzowsky, and G. Plunien, Phys. Rev. A {\bf 96}, 012512 (2017).

\bibitem{solovyev2019} D. Solovyev, T. Zalialiutdinov, A. Anikin, J. Triaskin, and L. Labzowsky, Phys. Rev. A {\bf 100}, 012506 (2019).

\bibitem{solovyevarxiv2019} D. Solovyev, arXiv:1905.13589v2 [physics.atom-ph] 28 Aug 2019

\bibitem{low} F. Low, Phys. Rev. {\bf 88}, 53 (1952).

\bibitem{olegreports} O. Yu. Andreev, L. Labzowsky, G. Plunien and D. Solovyev, {\it QED theory of the spectral line profile and its applications to atoms and ions}, Physics Reports {\bf 455} (2008).

\bibitem{physrep2018} T. A. Zalialiutdinov, D. A. Solovyev, L. N. Labzowsky and G. Plunien, {\it QED theory of multiphoton transitions in atoms and ions}, Physics Reports {\bf 737} (2018).

\bibitem{lab93} L. Labzowsky,  V. Karasiev,  I. Lindgren,  H. Persson,  S. Salomonson, Physica Scripta, {\bf 46}, pp. 150-156 (1993).

\bibitem{ambiguity} L. Labzowsky, D. Solovyev, and G. Plunien, Phys. Rev. A {\bf 80}, 062514 (2009).

\bibitem{akhiezer} A. I. Akhiezer and V. B. Berestetskii, {\it Quantum Electrodynamics}, Wiley, New York (1965).

\bibitem{shabaev2000} V. M. Shabaev, {\it Two-time Green function method in quantum electrodynamics of high-Z few-electron atoms}, Physics Reports {\bf 356} (2002).

\bibitem{dkb} V. M. Shabaev, I. I. Tupitsyn, V. A. Yerokhin, G. Plunien, and G. Soff, Phys. Rev. Lett. {\bf 93}, 130405 (2004).


\end{thebibliography}
\end{document}